\documentclass[12pt]{iopart}

\usepackage[T1]{fontenc}
\usepackage[latin1]{inputenc}

\expandafter\let\csname e\endcsname\relax
\expandafter\let\csname comment\endcsname\relax

\usepackage{setstack}
\usepackage{amssymb}

\usepackage{graphicx}

\newcommand{\braketop}[3]{{\langle{#1|{#2}|#3} \rangle}}
\newcommand{\bra}[1]{\langle{#1|}}
\newcommand{\ket}[1]{ {|#1}\rangle}
\newcommand{\e}{\epsilon}
\newcommand{\I}{^{-1}}
\newcommand{\f}[2]{\frac{#1}{#2} }

\newcommand{\comment}[1]{}
\newcommand{\rd}{\mathrm{d}}
\newcommand{\nn}{\nonumber}

\newcommand{\beginpmatrix}{\left( \begin{array}{ccc} }
\newcommand{\pmatrixend}{\end{array} \right)}

\newcommand{\beginBmatrix}{\left\{ \begin{array}{ccc} }
\newcommand{\Bmatrixend}{\end{array} \right\}}

\newcommand{\be}{\begin{eqnarray}}
\newcommand{\ee}{\end{eqnarray}}

\def\tl{\tilde}

\def\a{\alpha}
\def\g{\gamma}
\def\s{\sigma}

\def\n{\nu}

\def\g{\gamma}

\def\p{\phi}

\def\<{\langle}
\def\>{\rangle}
\def\*{\cdot}

\begin{document}

\title{Deriving identities for Wigner \{nj\}-symbols}

\author{ {\bf Gianluca Delfino} }

\address{School of Mathematical Sciences, University of Nottingham, Nottingham, NG7 2RD, UK.}
\ead{pmxgd2@nottingham.ac.uk}
\date{\small\today}

\begin{abstract}
We show how a simple and elegant graphical notation can be used to derive the Biedenharn-Elliott identity for the $6j$-symbol and we demonstrate how the same technique can be applied to obtain new identities for the $6j$. We then employ the same method also in the context of 4D spin-foam gravity and propose an analogous identity for the $15j$ symbol.

\end{abstract}


\section{INTRODUCTION}

The framework of spin foam quantum gravity provides a description for the discrete structure of space-time expressed in terms of variables of the representation theory of a certain gauge group (see \cite{Perez:2004hj} for introductory material). The 4D theory has undergone substantial development in the last fifteen years: since the seminal Barrett-Crane paper \cite{Barrett:1997gw}, a number of spin foam models have been suggested (for instance  \cite{Engle:2007uq} \cite{Freidel:2007py}). The 3D formulation of the theory, instead, has not change dramatically since the introduction of the very first spin foam model by Ponzano and Regge (PR) in 1968 \cite{Ponzano_Regge} (see \cite{Freidel:2004vi} for a thorough review). The model is now well understood and is known to reproduce the expected low energy behaviour \cite{Dowdall:2009eg}. 

Since the PR model is based on the $SU(2)$ group the amplitudes involve familiar objects of angular momentum theory. In particular, the most recurrent one is the $6j$-symbol, which represent the amplitude of the four-valent vertex of the theory. In spin-foam the vertex is the node of a graph dual to a triangulation (a spin-network) and represents a ``quanta'' of spacetime;  knowing the properties of the vertex means understanding the fundamental properties of the theory. In fact, the $6j$-symbol is known to satisfy some identities (see \cite{Schulten:1975yu} \cite{Bonzom:2009zd}) that are employed to prove triangulation independence of the PR model (see for instance \cite{Barrett:2008wh}). The purpose of this paper is to propose an elegant diagrammatic method able to generate and prove a wide class of such identities. Our method is powerful enough to encompass the common identities for the $6j$-symbol that were known from classical angular momentum theory, and also obtain new identities that, as far as the author is aware, were not present in the literature. Furthermore the method can be applied not only to the $6j$-symbol, but it can be employed to derive identities also for the $15j$ or any other $nj$ Wigner symbol.

It is of utmost importance that this diagrammatic formalism we propose can easily generate identities $within$ the framework of Spin-Foam quantum gravity. In fact, the key point of this paper is that method we introduce consists in carefully applying ``grasping'' operators on specific (flat) spin-networks (similar ideas have been pursued in \cite{Bonzom:2009zd}). To be precise, the grasping operator introduces an $extra$ link on the network connecting two edges, which do not necessarily belong to the same face. The action of the grasping then yields extra $6j$-symbols where the operator was applied. Finally, under certain conditions, moving the grasping operator along the spin-network edges produces equivalent geometrical configurations; the identity is obtained by simply equating the initial and final states. Choosing the spin-network determines which $nj$-symbol we are interested in, therefore the generated identities can be relevant for the study of both 3 and 4 dimensional spin foam models. 

These kind of identities carry a geometrical interpretation and are related to the Hamiltonian operator. In fact  the most common of these identities, the Biedenharn-Elliott (BE) identity, has been used by Freidel et al. as a discretization of the equation that imposes the scalar constraint in 3D spin-foam gravity \cite{Bonzom:2011hm}. In particular, the authors of \cite{Bonzom:2011hm} use the BE identity to impose flatness on a generic tetrahedron spin-network. 

Our work uses the paper \cite{Bonzom:2011hm} as a starting point. In section \ref{BE_id_section} we review the results of \cite{Bonzom:2011hm} and we show how the BE identity can be easily envisaged with a diagrammatic notation we propose. In section \ref{New_id_section}, instead, we  employ our diagrammatic formalism to derive new identities for the $6j$.  Finally and most importantly, in section \ref{4D_section}, we show that the same graphical method can be used to derive identities for the  $15j$-symbol, which are relevant in the context of the four-dimensional theory.

%
%
%
%

\subsection{3D Spin foam gravity and the Biedenharn-Elliott identity}

Spin foam models are defined by assigning probability amplitudes to elements of a triangulation. In general, given a triangulated space-time manifold, the spin foam model partition function will look like the following

\be\label{Generic_PF}
Z_{SF}= \sum_{j's} \prod_f A_f(j\ldots j)\prod_e A_e(j\ldots j)\prod_v A_v(j\ldots j),
\ee

where the labels denotes respectively $faces$, $edges$ and $vertices$ of the graph dual to the triangulation. The $j$'s are representation variables that depend on the group the spin-foam model is based on. For instance, in three dimensions, in the Ponzano-Regge model $A_v$ is the Wigner $6j$-symbol from angular momentum theory which is known to satisfy the Biedenharn-Elliott identity.

The aim of this section is not to introduce the reader to spin foam models (for which we refer to \cite{RovelliQG} \cite{Thiemann}), but rather to introduce the graphical notation we will extensively use in the subsequent sections. 

In section \ref{PR_section} we briefly review the construction of the Ponzano-Regge (PR) model, while in section \ref{BE_id_section} we derive the Biedenharn-Elliott identity following \cite{Bonzom:2011hm} and employing our graphical method.

\subsection{The Ponzano Regge Model}\label{PR_section}

Let us consider a three-dimensional manifold $M$ and principal $SU(2)$-bundle over it. Gravity then can be defined in terms of a cotriad $e_i^a$ and a connection $A^i_a$, where the indices $a,i=(1,2,3)$. Both $e$ and $A$ are 1-forms and the latter takes values in the $\mathfrak{su}(2)$ Lie algebra. The action for gravity in three dimension is then 

\be\label{BF_action}
S_{BF}[e,A]=\f{1}{4\pi G}\int_M \ e_i \wedge F^i(A),
\ee

where the 2-form $F^i(A)=\rd A^i+\f{1}{2}\e^{ijk}A^j\wedge A^k$ is the curvature of the connection $A$. Finally, if we define the density $E_i^a=\f{1}{2}\e^{abc}\e_{ijk}e^j_b e^k_c$ the phase space of gravity is spanned by the canonical pair 

\begin{equation*}
\{A^i_a(x), E_j^b(y) \}= \delta_a{}^b \delta^i{}_j \ \delta^{(2)}(x-y).
\end{equation*}

We can now discretise the theory by triangulating the manifold $M$. The fundamental ``block'' of spacetime is then the tetrahedron, and our variable $E_i^a$  has to be averaged over its edges. In fact, upon quantization, the canonical pair $(A^i_a,E_j^b)$ is replaced by $(g,X_e)\in SU(2)\times\mathfrak{su}(2)$, that are defined on the $dual$ graph to the triangulation. To be precise the group element $g$ is associated with the dual edge and defines the notion of $parallel$ $transport$ between to contiguous vertices; $X_e$ instead is the flux variable given by the smearing of the densitised triad $E^a$ 

\be
X_e=\int_e \ E_i (x)  \tau^i \rd x,
\ee 

where the $\tau^i$ are anti-Hermitian matrices\footnote{ Here $\tau_i=i\f{\s_i}{2}$, where $\s^i$ are the Pauli matrices.}.

Without delving into details we can state the PR partition function. Discretising the action in (\ref{BF_action}) we can derive:

\be\label{PR_partition_function}
Z_{PR}= \sum_j \ d_j\prod_v A_v(j\ldots j), \quad d_j=2j+1,
\ee

where the sum is over half-integers\footnote{The Ponzano-Regge model is based on $SU(2)$.} and, as mentioned, the $dual$ vertex amplitude is given by the $6j$-symbol known from angular momentum theory:

\begin{equation*}
A_v(j\ldots j)=\beginBmatrix j_1 &j_2 &j_3 \\j_4 &j_5 &j_6\Bmatrixend .
\end{equation*}

\begin{figure}[!ht]
\centering
\includegraphics[width=4cm]{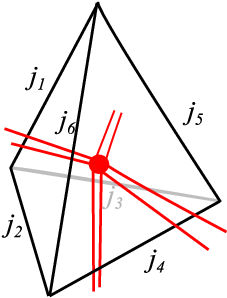}
\caption{\label{Tet_with_dual} The dual vertex in 3 dimensions.}
\end{figure}

\subsubsection{ Spin Networks}

Quantum geometries in spin foam are identified with spin-network states. These are simply dual graphs coloured (labelled) with representation variables and group elements. To be more precise, the edges of the triangulation carry a representation variable $j$, the dual edges $e$ are labelled by group elements $g$, finally invariant tensors $\iota$ sit on the dual vertices \cite{Baez:1997zt}. Spin networks span the kinematical Hilbert space of the theory and can be seen as elements of $L^2(SU(2)^e/SU(2)^v)$, i.e. functions over the group elements on the dual edges and invariant under translations acting at the dual vertices. Commonly a spin-networks is written in the from 

\be
s^{\{ j_e\}}(g1 \ldots g_n) = \  \bigotimes_{f} D^{(j)}(hol_f) \ \bigotimes_v \iota_v;
\ee
where $D^{(j)}(g)$ is the representation matrix of an element of the group $g\in G$. In this case $g=hol$ which stands for ``holonomy,'' i.e. the (ordered) product of all the group elements around a dual face $f$:  

\begin{equation*}
hol_f=\prod_{g_i\in \partial f} g_i.
\end{equation*}

\begin{figure}[!ht]
\centering
\includegraphics[width=8cm]{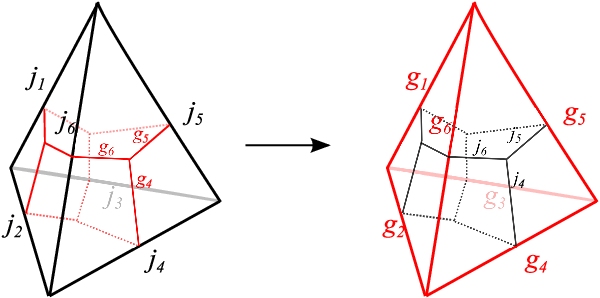}
\caption{\label{2d_Tet_with_dual} Two dimensional spin network and its dual.}
\end{figure}

 The simplest non trivial spin network one can envision in 2+1 dimensions is obtained by triangulating a $2$-sphere. The result is a tetrahedron which, in 2 dimensions, is in turn dual to a tetrahedron (see figure \ref{2d_Tet_with_dual}). 

Given the expression for the $\{6j\}$ in (\ref{6j_Symbol}) we can easily write the spin-network state of the tetrahedron in figure \ref{2d_Tet_with_dual}:

\begin{eqnarray}
\fl \nn  s^{\{j_e\}}_{\rm Tet}(g_1,\ldots,g_6) = &   (-1)^{j_2-a_2+j_3-a_3} \ (-1)^{j_4-a_4} \ (-1)^{j_1-a_1+j_5-a_5+j_6-a_6}  \  \prod_i^6 \braketop{j_i,a_i}{g_i}{j_i,b_i}  \times\\
 \fl \nn & \beginpmatrix j_1 & j_2 & j_3\\ -a_1 &-a_2 &-a_3\pmatrixend  \beginpmatrix j_3 &j_4 &j_5\\ b_3 & b_4 & b_5\pmatrixend  \beginpmatrix j_4 &j_2 &j_6\\-a_4 & b_2 & b_6\pmatrixend  \beginpmatrix j_1 &j_5 &j_6\\b_1 &-a_5 & -a_6\pmatrixend, \\ \label{Tet_spin_network}
\end{eqnarray}

where the $(3jm)$-symbols are the invariant tensors $\iota_v$ sitting on each vertex (see appendix \ref{RecouplingTheoryAppendix}) and the $D^j_{mn}(g)=\braketop{j,m}{g}{j,n}$ are representation matrices of the group elements in their respective representations.

The key point is that in 3D, as can be seen at the classical level from the equations of motion of (\ref{BF_action}), the physical states are $flat$ geometries. This translates, in terms of quantum geometries, to imposing the holonomy $g_1\ldots g_n$ around dual faces to be trivial. Therefore, concerning the spin network  (i.e. the quantum geometry) in (\ref{Tet_spin_network}),  the $physical$ configuration is given by contracting with a series of Dirac deltas on the group. In fact, if we take

\be\label{phys_tetrahedron}
\psi_{ phys}(g_1,\ldots,g_6) = \delta(g_4 g_5 g_6)\,\delta(g_1 g_6 g_2\I)\,\delta(g_2 g_4 g_3\I).
\ee

we can obtain the {6j}-symbol by taking the inner product of the spin network function $s^{\{j_e\}}_{\rm tet}(g_1,\ldots,g_6)$:

\begin{eqnarray}\label{Integral_with_deltas}
\nn \psi_{phys}(j_1,\ldots,j_6) &= \int \prod_{e=1}^6 dg_e\ s^{\{j_e\}}_{\rm tet}(g_1,\ldots,g_6)\ \psi_{phys}(g_1,\ldots,g_6),\\
&= \beginBmatrix j_1 &j_2 &j_3 \\j_4 &j_5 &j_6 \Bmatrixend.
\end{eqnarray}

\subsection{The Biedenharn-Elliott identity}\label{BE_id_section}

The  authors of  \cite{Bonzom:2011hm} work in the context of $(2+1)$ spin-foam quantum gravity. They introduce a quantum $Hamiltonian$ constraint and prove that the BE identity can be seen a consequence of imposing such constraint on a flat tetrahedron.

Following \cite{Bonzom:2011hm}, define a quantum constraint $\hat{H}$ that acts on spin-networks $\ket{s}$. The idea is to choose two contiguous fluxes variables $X_i,X_j\in \mathfrak{su}(2)$, and impose that their scalar product is invariant under parallel transport around the dual face:

\be\label{H_equation}
\hat{H}_{ij} \ket{s}=0, \quad where \quad \hat{H}_{ij}= X_i\cdot X_{j}-X_i\cdot Ad_{hol} X_{j}.
\ee

Here the flux $X_j$ is parallel-transported with the holonomy through the action of $G$ on its Lie algebra: $g \triangleright X= gXg\I=Ad_g X.$ We note that the mere action of $\hat{H}_{ij}$ on a node of the spin-network does not impose flatness, it implies only the fact that the holonomy $g$ lives in the Cartan subalgebra spanned by the Lie-algebra element $X_j$. However imposing the constraint (\ref{H_equation}) at each node $does$ imply that the holonomy is trivial. In other words, if the constraint is applied to each node, this imposes $flatness$ of the holonomies around all the faces (see equation \ref{phys_tetrahedron}). 

One could take the opposite point of view. Assuming flatness of the holonomy (again see equation \ref{phys_tetrahedron}), we notice that the constraint in (\ref{H_equation}) is trivially satisfied. The next step is to show that such condition, when $\ket{s}$ is the tetrahedron spin-network in (\ref{Tet_spin_network}), implies a second order recursion relation on the $6j$-symbol \cite{Bonzom:2011hm}. This recursion relation is commonly known in Angular Momentum theory as the Biedenharn-Elliott (BE) identity and it reads

\begin{eqnarray} \label{Classical_BE_id}
\fl \nn A_{+1}(j_1)\,\beginBmatrix j_1+1 &j_2 &j_3 \\ j_4 &j_5 &j_6 \Bmatrixend + A_{0}(j_1)\,\beginBmatrix j_1 &j_2 &j_3 \\ j_4 &j_5 &j_6 \Bmatrixend + A_{-1}(j_1)\ \beginBmatrix j_1-1 &j_2 &j_3 \\ j_4 &j_5 &j_6 \Bmatrixend = 0.\\ 
\end{eqnarray}

The coefficients are given by

\begin{eqnarray} \label{Coefficient_A_0}
\fl \nn A_0(j_1) = (-1)^{j_2+j_4+j_6}\beginBmatrix j_2 & j_2 &1 \\ j_6 &j_6 &j_4\Bmatrixend + \\    +(-1)^{2j_1+j_2+ j_3+j_5+j_6}(2j_1+1)\ \beginBmatrix j_1 & j_1 &1\\ j_2 &j_2 &j_3\Bmatrixend\,\beginBmatrix j_1 & j_1 &1\\ j_6 &j_6 &j_5\Bmatrixend,
\end{eqnarray}

and

\begin{eqnarray}\label{Coefficient_Apm1}
\fl \nn A_{\pm 1}(j_1)  = (-1)^{2j_1+j_2+ j_3+j_5+j_6+1}\bigg(2(j_1\pm1)+1\bigg)\ \beginBmatrix j_1\pm1 & j_1 &1\\ j_2 &j_2 &j_3\Bmatrixend\,\beginBmatrix j_1\pm 1 & j_1 &1\\ j_6 &j_6 &j_5\Bmatrixend. \\
\end{eqnarray}


In the following we show how to prove the BE identity the same way it is proved in \cite{Bonzom:2011hm}. In this section we will also introduce the graphical notation we employed to find our results. We show that with such graphical notation the derivation of the result is greatly simplified.

We begin by choosing the contiguous fluxes to be $X_2$ and $X_6$ (see figure \ref{2d_Tet_with_dual}). Then we can write the equation for $\hat{H}$ action on $s^{\{j_e\}}_{\rm Tet}$ (to avoid cluttering of the notation we dropped the $ket$ around the spin-network state) as

\begin{equation}\label{BEid}
  (X_2 \cdot X_6)\ s^{\{j_e\}}_{\rm Tet}(g_1,\ldots,g_6)=  (X_2 \cdot Ad_{g_2\I  g_1 g_6\I} X_6)\ s^{\{j_e\}}_{\rm Tet}(g_1,\ldots,g_6)
\end{equation}

where the $X_e$'s act by inserting a $\tau_e$ in front of the corresponding ket $\ket{j_e,a_e}$ in (\ref{Tet_spin_network}). The right hand side (rhs) is given by taking the same product $after$ we parallel transported $X_6$ around the dual face as in

\begin{equation*}
X_6\rightarrow Ad_{g_2\I  g_1 g_6\I}  \ X_6 .
\end{equation*}

\begin{figure}[!ht]
\centering
\includegraphics[width=10cm]{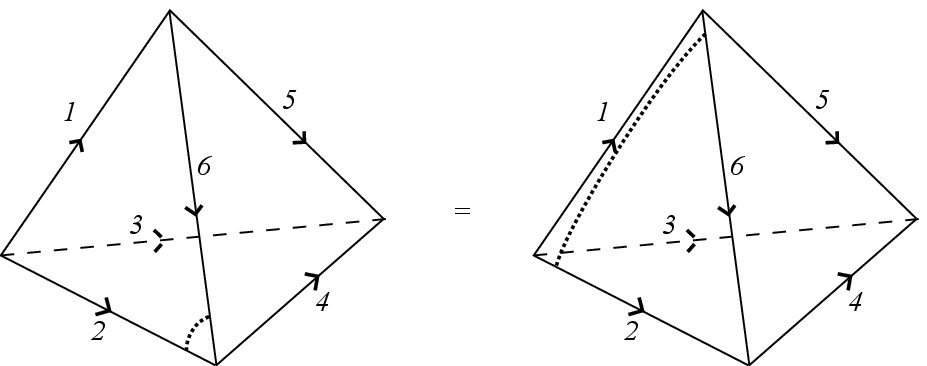}
\caption{\label{BEidFig}Biedenharn-Elliott identity.}
\end{figure}

The key point of our analysis is that relation (\ref{BEid}) (and $any$ other of this kind) can be represented graphically as inserting ``grasping'' operators in different places of the spin network. In this case (\ref{BEid}) is simply given by the figure \ref{BEidFig}, where the grasping operator is represented by a dashed line. We notice that the grasping operator has been moved along the lines $2$ and $6$.

In the following section we will give the details of how to prove the BE identity explaining every step diagrammatically. We split the computation in left and right hand sides.

\subsubsection{LHS of BE identity proof}

\begin{figure}[!ht]
\centering
\includegraphics[width=10cm]{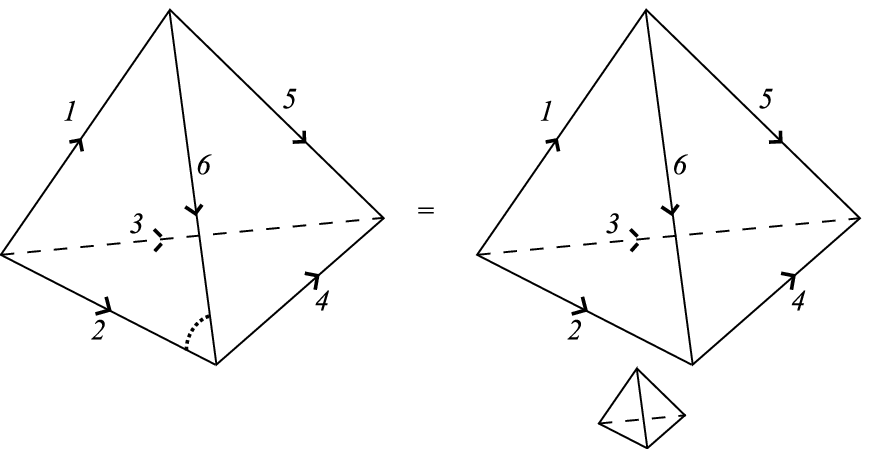}
\caption{\label{BEstepsA} Diagrammatic  computation of LHS.}
\end{figure}

To obtain the BE identity we first evaluate the LHS of (\ref{BEid}). Graphically, figure \ref{BEstepsA} shows how we want to proceed. We can already see that this side of the equation is rather trivial in the graphical notation.

What exactly happens is that the ``grasping'' in fig \ref{BEstepsA} shows us where the operator $X_2\cdot X_6$ is acting. Then we can use recoupling theory to $extract$ a $6j$-symbol which is represented by a small tetrahedron (for convenience of the reader we listed some useful recoupling theory relations in the appendix \ref{RecouplingTheoryAppendix}).

In formulas the action of the grasping operator is given by the insertion of $-(\tau_2){}^n \otimes (\tau_6 ){}_n$\footnote{Since we are using anti-Hermitian generators $X\cdot X=-X^i X_i$} in the spin network.

Following \cite{Bonzom:2011hm} we adopt the relations

\begin{equation}
(\tau_e )^n\otimes (\tau_{e^\prime})_n = \sum_{n=-1}^{+1}\, (-1)^{1-n}\ (L_e )^n\otimes (L_{e^\prime} )_{-n},
\end{equation}
where we defined $L_0=-\f{\s_z}{2}$ and $L_\pm=\f{(\pm \s_x+i\s_y)}{2\sqrt{2}}$.

To avoid further cluttering of the notation, in the following we will omit the subscript $e$ that indicates which edge the operators act on.
 
The action of the $L_n$ operators on $SU(2)$ kets is easily defined using $\{3jm\}$ symbols:

\be\label{LOnKet}
\nn L_n\ket{j,a}=&-N_j \sum_b (-1)^{j-b} \ \beginpmatrix 1 & j & j\\ n & a & -b\pmatrixend \ket{j,b}=\\ &(-1)^{2j+1} N_j \sum_b  (-1)^{1+j-n-a} \ \beginpmatrix 1 & j & j\\ -n & -a & b\pmatrixend \ket{j,b}, 
\ee
where the normalization factor is 

\begin{equation}\label{normFactor}
  N_j=\sqrt{j(j+1)d_j}, \quad d_j=2j+1.
\end{equation}

We can now see how, using recoupling theory, we obtain the desired result on the LHS. Writing only  relevant parts of the spin-network state $s^{\{j_e\}}_{\rm Tet}$, we have: 

\begin{eqnarray}
\fl \nn (X_2 \cdot X_6)\  \sum_{a,b} \ (-1)^{j_4-a_4} \beginpmatrix j_4 &j_2 &j_6\\-a_4 & b_2 & b_6\pmatrixend  \  \ket{j_2,b_2}\otimes \ket{j_6,b_6}=
\\ 
\fl\nn -\ \sum_{a,b} (-1)^{j_4-a_4} \beginpmatrix j_4 &j_2 &j_6\\-a_4 & b_2 & b_6\pmatrixend  \ \tau_{i} \ket{j_2,b_2}\otimes \tau^{i}\ket{j_6,b_6}=
\\
\fl\nn -\ \sum_{n=-1}^{+1}\, (-1)^{1-n}\    \sum_{a,b} (-1)^{j_4-a_4} \beginpmatrix j_4 &j_2 &j_6\\-a_4 & b_2 & b_6\pmatrixend  \ L^n \ket{j_2,b_2}\otimes L_{-n}\ket{j_6,b_6}=
\\ \comment{
\nn - (-1)^{2j_2+1} (-1)^{2j_6+1}\ N_{j_2} N_{j_6} \ \sum_{n=-1}^{+1}\, (-1)^{1-n}\ \sum_{b_2} (-1)^{j-b_2}   \ \beginpmatrix 1 & j_2 & j_2\\ -n & -b_2 & k_2\pmatrixend \ \sum_{b_6} (-1)^{j-b_6} \ \beginpmatrix 1 & j_6 & j_6\\ n & -b_6 & k_6\pmatrixend   
\\ \nn   \sum_{a,b} (-1)^{j_4-a_4} \beginpmatrix j_4 &j_2 &j_6\\-a_4 & b_2 & b_6\pmatrixend  \ \ket{j_2,k_2}\otimes \ket{j_6,k_6}=
\\ \comment{I flip the j6 one k6 so i got another (-)^{2j6+1}, and i flipped the line with b2 so i got an extra (-)^{2j}}
\nn - (-1)^{2j_2+1} \ N_{j_2} N_{j_6} \ \sum_{n=-1}^{+1}\, (-1)^{1-n}\ \sum_{b_2} (-1)^{j-b_2}   \ \beginpmatrix 1 & j_2 & j_2\\ -n & -b_2 & k_2\pmatrixend \ \sum_{b_6} (-1)^{j-b_6} \ \beginpmatrix 1 & j_6 & j_6\\ n & k_6 & -b_6\pmatrixend   
\\ \nn  \sum_{a,b} (-1)^{j_4-a_4} \beginpmatrix j_4 &j_2 &j_6\\-a_4 & b_2 & b_6\pmatrixend  \ \ket{j_2,k_2}\otimes \ket{j_6,k_6}=
\\
}\fl \nn \ N_{j_2} N_{j_6} (-)^{j_2+ j_4+j_6} \beginBmatrix j_2 &j_2 &1 \\j_6 &j_6 &j_4\Bmatrixend \sum_{a,b}  (-1)^{j_4-a_4}  \beginpmatrix j_4 &j_2 &j_6\\ -a_4 & k_2  &k_6\pmatrixend  \ket{j_2,k_2}\otimes \ket{j_6,k_6}. \\
\end{eqnarray}
 
In the last step we used the relations (\ref{flipping}) and (\ref{grasping}). Therefore the overall effect of the insertion of the grasping operator on the left-hand-side of the equation in (\ref{BEid}) is simply:

\begin{equation}\label{LHS}
\fl  (X_2 \cdot X_6)\ s^{\{j_e\}}_{\rm Tet}(g_1,\ldots,g_6)=  \ N_{j_2} N_{j_6} (-)^{j_2+ j_4+j_6} \beginBmatrix j_2 &j_2 &1 \\j_6 &j_6 &j_4\Bmatrixend \ s^{\{j_e\}}_{\rm Tet}(g_1,\ldots,g_6)
\end{equation}

 \subsubsection{RHS  of BE identity proof}

Before proceeding with the evaluation of the right hand side of (\ref{BEid}) we define $\tl{X}=g\I X g$, so that

\begin{equation}\label{rhs_BEid}
\fl    (X_2 \cdot Ad_{g_2\I  g_1 g_6\I} X_6)\ s^{\{j_e\}}_{\rm Tet}(g_1,\ldots,g_6)= (\tl{X_2} \cdot Ad_{g_1} \tl{X_6}) \ s^{\{j_e\}}_{\rm Tet}(g_1,\ldots,g_6).
\end{equation}

This way the evaluation of the action of the quantum operator is more straightforward. However we note that the insertion of $g\I X g$ in front of the respective kets reduces to $X$ effectively acting on the bra:

\be
\braketop{j,a}{g \;g\I X g }{j,b}=\braketop{j,a}{X \; g}{j,b}.
\ee

The operator $Ad(g)$ acts as multiplicative operator, contributing adding representation matrices in the adjoint (spin $1$) representation as factors:

\begin{eqnarray*}
Ad(g) \rightarrow  \ D^1(g)_{m n}.
\end{eqnarray*}

Analogously to (\ref{LOnKet}) we have

\begin{eqnarray}\label{LOnBra}
\fl \nn \langle j_2, a_2\vert L_{m} = (-1)^{2j_2+1} N_{j_2}\sum_{k_2} (-1)^{1-m} (-1)^{j_2-k_2}\beginpmatrix 1 &j_2 &j_2\\-m &-k_2 &a_2\pmatrixend\,\bra{j_2, k_2},\\
\fl  \langle j_6, a_6\vert (-1)^{1-n}L_{-n} = (-1)^{2j_6+1} N_{j_6}\sum_{k_6}  (-1)^{j_6-k_6}\beginpmatrix 1 &j_6 &j_6\\n &-k_6 &a_6\pmatrixend\,\langle j_6, k_6\vert.
\end{eqnarray}

with the same normalization factors (\ref{normFactor}).

\begin{figure}[!ht]
\includegraphics[width=15cm]{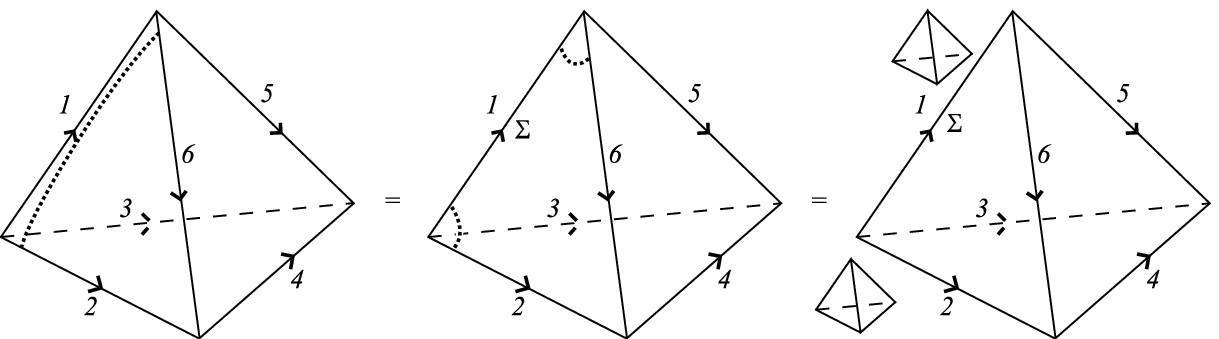}
\caption{Diagrammatic computation of RHS.\label{Fig5}}
\end{figure}

The operation to carry out can be also easily described with the formalism that we are proposing. The steps are in figure \ref{Fig5}.

Precisely, each operator $\tl{X}_e^m$ acts by inserting a $\tau_{e}^m$ in front of the respective bra $\bra{j_e,a_e}$, while $Ad_{g_e}$ acts as a multiplicative operator in the adjoint representation as $\braketop{1,m}{g_1}{1,n}$.  We notice that the insertion of the grasping operators, after employing recoupling theory,  yields two extra $\{6j\}$'s:

\begin{eqnarray}
\fl \nn (\tl{X}_2 \cdot Ad_{g_1} \tl{X}_6 )  \ (-1)^{j_2-a_2+j_3-a_3} \ (-1)^{j_1-a_1+j_6-a_6} \beginpmatrix j_1 & j_2 & j_3\\ -a_1 &-a_2 &-a_3\pmatrixend  \beginpmatrix j_1 &j_5 &j_6\\b_1 &-a_5 & -a_6\pmatrixend \times
\\
\nn \braketop{j_1,a_1}{g_1}{j_1,b_1}  \bra{j_2,a_2}g_2  \otimes \bra{j_6,a_6}g_6= 
\\
\fl \nn (-)\sum  \ \braketop{1,m}{g_1}{1,n}\braketop{j_1,a_1}{g_1}{j_1,b_1}   \ (-1)^{j_2-a_2+j_3-a_3} \ (-1)^{j_1-a_1+j_6-a_6} 
  \beginpmatrix j_1 & j_2 & j_3\\ -a_1 &-a_2 &-a_3\pmatrixend 
\\
\fl \nn \beginpmatrix j_1 &j_5 &j_6\\b_1 &-a_5 & -a_6\pmatrixend  \bra{j_2,a_2}L_{m}  g_2  \otimes \bra{j_6,a_6} L_{-n} (-1)^{1-n} g_6,
\end{eqnarray}
which after using the recoupling relation (\ref{parallel_momenta}) becomes:
\begin{eqnarray}
 \nn   N_{j_2}   N_{j_6} \ \sum_{J_1=j_1-1}^{j_1+1} d_{J_1} \  \ (-1)^{J_1+j_1 +j_2+j_3+j_5+j_6+1}\ \beginBmatrix 1 & J_1 & j_1\\ j_3 & j_2 & j_2\Bmatrixend \ \beginBmatrix 1 & J_1 & j_1\\ j_5 & j_6 & j_6\Bmatrixend \times
\\
\nn  (-1)^{j_2-k_2+j_3-a_3} \ (-1)^{J_1-A_1+j_6-k_6} \beginpmatrix J_1 & j_2 & j_3\\ -A_1 & -k_2 & -a_3\pmatrixend \ \beginpmatrix J_1 & j_5 & j_6\\ B_1 & -a_5 & -k_6\pmatrixend \times \\  \braketop{J_1,A_1}{g_1}{J_1,B_1} \  \ \bra{j_2.k_2}g_2\otimes \bra{j_6, k_6}g_6.
\end{eqnarray}


Finally  we obtain

\be\label{RHS}
\fl \nn (\tl{X}_2 \cdot Ad_{g_1} \tl{X}_6 )  s^{\{j_e\}}_{\rm Tet}(g)=
\\
\nn  N_{j_2}   N_{j_6}  \sum_{J_1=j_1-1}^{j_1+1} d_{J_1}  (-1)^{J_1+j_1 +j_2+j_3+j_5+j_6+1} \beginBmatrix 1 & J_1 & j_1\\ j_3 & j_2 & j_2\Bmatrixend  \beginBmatrix 1 & J_1 & j_1\\ j_5 & j_6 & j_6\Bmatrixend  s^{\{J_1\ldots j_e\}}_{\rm Tet}(g).\\
\ee

\subsubsection{The BE identity}

Putting together the results in (\ref{LHS}) and in (\ref{RHS}) we have:

\be
\fl  \hat{H}\ s^{\{j_e\}}_{\rm Tet}=  N_{j_2} N_{j_6} \left[ A_0(j) \ s^{\{j_e\}}_{\rm Tet}  +  A_{-1}(j) \ s^{\{j_1-1,j_2 \ldots j_6\}}_{\rm Tet}+A_{+1}(j) s^{\{j_1+1,j_2 \ldots j_6\}}_{\rm Tet} \right]=0. 
\ee
The coefficients $A_i(j)$ are exactly the ones in (\ref{Coefficient_A_0}). 

Now the last step. We evaluated the action of the quantum operator $\hat{H}$ on a specific spin-network $s^{\{j_e\}}_{\rm Tet}$, therefore we know its action also on a linear superposition of such states like $ \Psi(g)= \sum  \ \prod  d_j  \ \psi(j) \  s_{\rm Tet}(g)$:

\be\label{BE_on_sup_ofStates}
\fl \nn \hat{H} \ \Psi(g_1 \ldots g_6)= \sum_{j_1 \ldots j_6} \left[ \prod_{i=1}^6 d_j  \right]  \psi(j_1 \ldots j_6)  \hat{H} \  s^{\{j_e\}}_{\rm Tet}(g_1,\ldots,g_6)
\\
\nn = \sum_{j_1 \ldots j_6} \left[ \prod_{i=2}^6 d_j  \right]     N_{j_2} N_{j_6}   \left[ d_{j_1}  A_0(j) \  \psi(j_1 \ldots j_6) \right. \\ \nn \left. + d_{j_1+1} \psi(j_1+1 \ldots j_6) A_{-1}(j+1)+d_{j_1-1} \psi(j_1-1 \ldots j_6) A_{+1}(j-1) \right]  s^{\{j_e\}}_{\rm Tet}=0. \\
\ee 

Finally we note that the following relations hold for the coefficients 
\be\label{dA=dA}
d_{j\mp1}A_{\pm}(j\mp1)=d_jA_{\mp}(j).
\ee

 Thus we obtain that the operator $\hat{H}$ imposes

\be 
\fl  A_{0}(j)\psi(j_1 \ldots j_6) + A_{+1}(j)\,\psi(j_1+1 \ldots j_6) + A_{-1}(j)\psi(j_1-1 \ldots j_6) = 0,
\ee

which is exactly the Biedenharn-Elliott identity in (\ref{Classical_BE_id}).

\section{New Identity}\label{New_id_section}

In the previous section we showed how we can define an operator $\hat{H}$ to impose flatness on the dual faces of a spin network. Most importantly we showed how, assuming flatness of the tetrahedron, the equation in (\ref{H_equation}) implies the BE identity on the $6j$-symbol, which is used in spin foam to prove triangulation independence of PR model and to study the ``low energy'' limit of the theory. 

We saw that the action of the operator $\hat{H}$ on the spin networks can be seen as the insertion of a grasping operators (represented with a dashed line in figure \ref{BEidFig}) moved along the edges of the spin-network. 

In this section we want find a new recursion identity for the $\{6j\}$ by adopting the same principles. However we will insert the grasping operator to connect two edges that belong to $different$ dual faces and move the grasping along these edges. For convenience of the reader we state the new identity before delving into the details of the derivation. The new identity for the $\{6j\}$ derived with our graphical method reads

\begin{eqnarray}\label{Final_NewId}
\fl \nn - A1_0(j_1) \psi(j_1 \ldots j_6) -   A1_{+1}(j_1) \psi(j_1+1 \ldots j_6)-  A1_{-1}(j_1)\psi(j_1-1 \ldots j_6)  
\\
\nn +A4_0(j_4) \psi(j_1 \ldots j_6) +   A4_{+1}(j_4)\psi(j_4+1 \ldots j_6) + A4_{-1}(j_4)\psi( j_4-1 \ldots j_6)=0,\\
\end{eqnarray}

Where the coefficients are:
\begin{eqnarray}\label{NewId_coefficients}
\nn &A1_0(j)=(-1)^{2j_1} d_{j_1} \beginBmatrix 1 & j_1 & j_1\\ j_3 & j_2 & j_2\Bmatrixend \ \beginBmatrix 1 & j_1 & j_1\\ j_6 & j_5 & j_5\Bmatrixend 
\\
\nn &A1_{\pm 1}(j)=(-1)^{2j_1} d_{j_1\pm1} \beginBmatrix 1 & j_1\pm1 & j_1\\ j_3 & j_2 & j_2\Bmatrixend \ \beginBmatrix 1 & j_1\pm1 & j_1\\ j_6 & j_5 & j_5\Bmatrixend 
\\ \comment{
\nn &A1_{+1}(j)=(-1)^{2j_1} d_{j_1+1} \beginBmatrix 1 & j_1+1 & j_1\\ j_3 & j_2 & j_2\Bmatrixend \ \beginBmatrix 1 & j_1+1 & j_1\\ j_6 & j_5 & j_5\Bmatrixend 
\\ }
\nn &A4_0(j)= (-1)^{2j_4} d_{j_4} \beginBmatrix 1 & j_4 & j_4\\ j_3 & j_5 & j_5\Bmatrixend  \  \beginBmatrix 1 &j_4 & j_4\\ j_6 & j_2 & j_2\Bmatrixend 
\\
\nn &A4_{\pm1}(j)= (-1)^{2j_4} d_{j_4\pm1} \beginBmatrix 1 & j_4\pm1 & j_4\\ j_3 & j_5 & j_5\Bmatrixend  \  \beginBmatrix 1 &j_4\pm1 & j_4\\ j_6 & j_2 & j_2\Bmatrixend \comment{
\\ 
\nn &A4_{+1}(j)= (-1)^{2j_4} d_{j_4+1} \beginBmatrix 1 & j_4+1 & j_4\\ j_3 & j_5 & j_5\Bmatrixend  \  \beginBmatrix 1 &j_4+1 & j_4\\ j_6 & j_2 & j_2\Bmatrixend }.\\
\end{eqnarray}

\begin{center}
\begin{figure}[!ht]
\centering
\includegraphics[width=10cm]{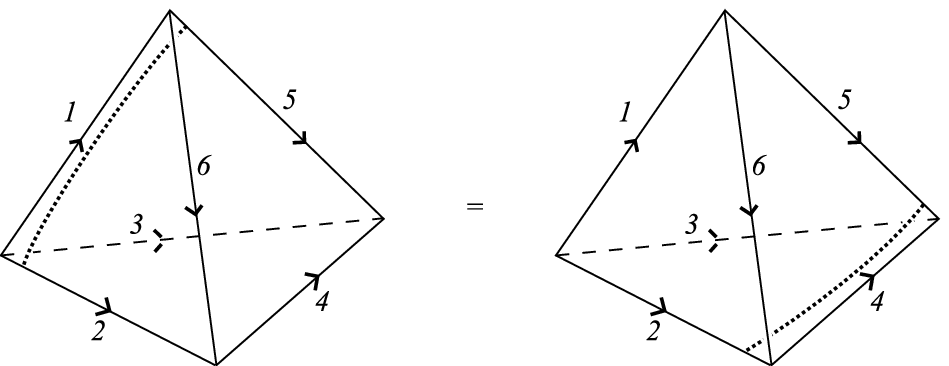}
\caption{\label{NewIdfig} New proposed Identity.}
\end{figure}
\end{center}

In the following we give a detailed derivation for the new identity in equation \ref{Final_NewId} using the method introduced in the previous section. In the graphical method, the new starting equation is represented in figure \ref{NewIdfig}. In this case we will use the fact that imposing flatness on the dual faces implies that also the holonomies around $edges$ are trivial: looking at figure \ref{2d_Tet_with_dual}, imposing the deltas in (\ref{phys_tetrahedron}), then also $g_1g_2g_4g_5=1.$

Thus we can define a new $\hat{H}_{new}=  (\tl{X_2} \cdot Ad_{g_1} \tl{X_5}) -(X_2 \cdot Ad_{g_4} X_5) $ operator which yields the identity:
\begin{equation}\label{NewId}
  (\tl{X_2} \cdot Ad_{g_1} \tl{X_5}) \ s^{\{j_e\}}_{\rm Tet}(g_1,\ldots,g_6)= (X_2 \cdot Ad_{g_4} X_5) \ s^{\{j_e\}}_{\rm Tet}(g_1,\ldots,g_6).
\end{equation}

\begin{figure}[!ht]
\centering
\includegraphics[width=12cm]{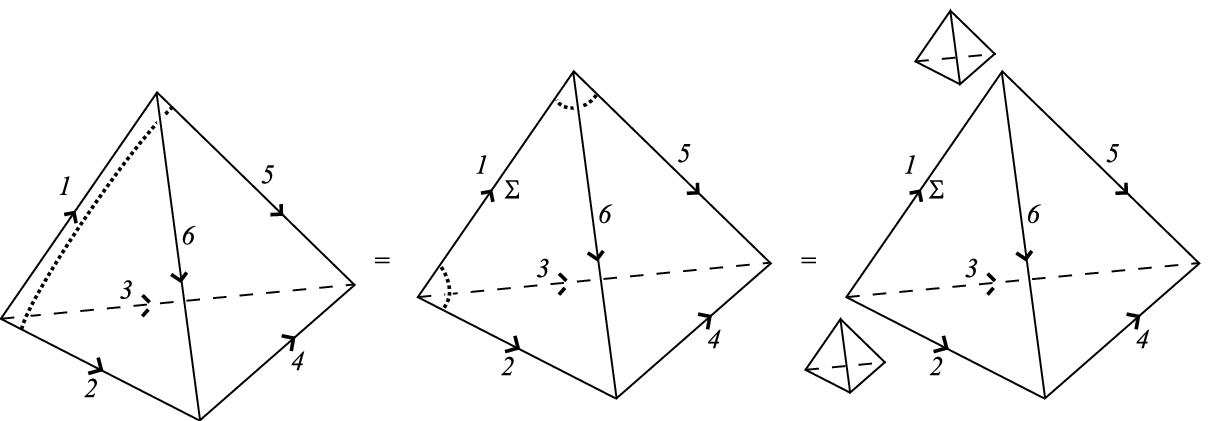}
\caption{\label{NewIdStepsAfig} Steps for the LHS}
\end{figure}

Let us split the calculation into two parts and evaluate both sides. Graphically the calculation we want to follow for the  left hand side of (\ref{NewId})  goes as in figure \ref{NewIdStepsAfig}.

Again the dashed lines represent grasping operators. In the second step we used, as before, recoupling theory rule (\ref{grasping}).

As in section \ref{BE_id_section}, Each operator $\tl{X}_e^m$ acts by inserting a $\tau_{e}^m$ in front of the respective bra $\bra{j_e,a_e}$, while $Ad_{g_e}$ acts as a multiplicative operator in the adjoint representation as $D_{mn}^1(g)$. Explicitly it reads

\begin{eqnarray}
\fl \nn (\tl{X_2} \cdot Ad_{g_1} \tl{X_5}) \ s^{\{j_e\}}_{\rm Tet}(g)= 
\\ 
\nn N_{j_2}   N_{j_5}  \sum_{J_1=j_1-1}^{j_1+1} d_{J_1}  (-1)^{2J_1+j_2+j_3 +j_5+j_6+1} \beginBmatrix 1 & J_1 & j_1\\ j_3 & j_2 & j_2\Bmatrixend \beginBmatrix 1 & J_1 & j_1\\ j_6 & j_5 & j_5\Bmatrixend \ s^{\{J_1,\ldots j_e \}}_{\rm Tet}(g),\\
\end{eqnarray}


where $N_j$ is the same normalization factor in (\ref{normFactor})

\begin{figure}[!ht]
\centering
\includegraphics[width=12cm]{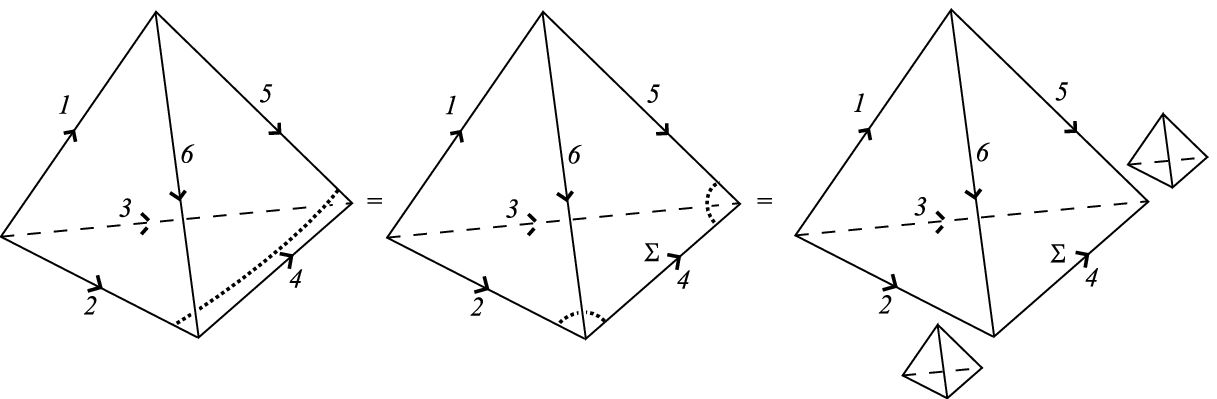}
\caption{\label{NewIdStepsBfig} Steps for the RHS.}
\end{figure}

On the right hand side we have instead (see figure \ref{NewIdStepsBfig})

\begin{eqnarray}
\fl \nn  (X_2 \cdot Ad_{g_4} X_5) s^{\{j_e\}}_{\rm Tet}(g) = \comment{  \\
&    (-1)^{j_4-a_4} \ \beginpmatrix j_3 &j_4 &j_5\\ b_3 & b_4 & b_5\pmatrixend  \beginpmatrix j_4 &j_2 &j_6\\-a_4 & b_2 & b_6\pmatrixend  \\
 			&\braketop{j_4,a_4}{g_4}{j_4,b_4} g_2\ket{j_2,b_2} \otimes g_5\ket{j_5,b_5}=
\\}
\\ \nn  N_{j_2}  N_{j_5}  \sum_{J_4=j_4-1}^{j_4+1} d_{J_4}  (-1)^{2J_4+j_2+j_3 +j_5+j_6+1}\beginBmatrix 1 & J_4 & j_4\\ j_3 & j_5 & j_5\Bmatrixend    \beginBmatrix 1 &J_4 & j_3\\ j_6 & j_2 & j_2\Bmatrixend \ s^{\{J_4\ldots j_e\}}_{\rm Tet}(g).\\
\end{eqnarray}

All together:

\begin{eqnarray}
\fl \nn   N_{j_2}   N_{j_5} \ \sum_{J_1=j_1-1}^{j_1+1} d_{J_1} \  \ (-1)^{2J_1+j_2+j_3 +j_5+j_6+1}\ \beginBmatrix 1 & J_1 & j_1\\ j_3 & j_2 & j_2\Bmatrixend \ \beginBmatrix 1 & J_1 & j_1\\ j_6 & j_5 & j_5\Bmatrixend  \ s^{\{J_1\ldots j_e\}}_{\rm Tet}-
\\
\fl \nn  N_{j_2}  N_{j_5}\  \sum_{J_4=j_4-1}^{j_4+1} d_{J_4} \ (-1)^{2J_4+j_2+j_3 +j_5+j_6+1}\ \beginBmatrix 1 & J_4 & j_4\\ j_3 & j_5 & j_5\Bmatrixend  \  \beginBmatrix 1 &J_4 & j_3\\ j_6 & j_2 & j_2\Bmatrixend \ s^{\{J_4\ldots j_e\}}_{\rm Tet}=0. \\
\end{eqnarray}

As in equation (\ref{BE_on_sup_ofStates}),  we can use the result to evaluate the action of $\hat{H}_{new}$ on a superposition of states. We get
\begin{eqnarray}
\fl \nn  \hat{H}_{new} \ \Psi(g_1 \ldots g_6)
= N_{j_2} N_{j_5} \sum_{j_1 \ldots j_6}  \prod_{i=2,3,5,6}  d_{j_i} \times
\\
\fl \nn    \bigg[ d_{j_4} \bigg( d_{j_1}  A1_0(j_1) \psi(j_e) + d_{j_1+1}  A1_{-1}(j_1+1) \psi(j_1+1)+d_{j_1-1}  A1_{+1}(j_1-1)\psi(j_1-1)\bigg)  
\\
\fl \nn   - d_{j_1}\bigg( d_{j_4}  A4_0(j_4) \psi(j_4) + d_{j_4+1}  A4_{-1}(j_4+1)\psi( j_4+1) +d_{j_4-1}  A4_{+1}(j_4-1)\psi( j_4-1)\bigg)  \bigg] \ s^{\{j_e\}}_{\rm Tet}=0,\\
\end{eqnarray}

where the coefficients are the one in (\ref{NewId_coefficients}).

Using again the relation (\ref{dA=dA}) we \comment{ have

\be
 \sum_{j_1 \ldots j_6}  \prod_i^6   d_{j_i}  N_{j_2} N_{j_5} \left[  A1_0(j_1) \psi(j_e) +   A1_{+1}(j_1) \psi(j_1+1)+  A1_{-1}(j_1)\psi(j_1-1)  \right.
\\
\left.  A4_0(j_4) \psi(j_4) +   A4_{+1}(j_4)\psi( j_4+1) + A4_{_1}(j_4)\psi( j_4-1)  \right] \ s^{\{j_e\}}_{\rm Tet}=0 .
\ee 

Thus we }obtain the identity for the $6j$-symbol in (\ref{Final_NewId}).

We want to stress that the derivation of this new identity is itself a proof of it. In fact, starting from the assumption of flatness of the holonomy $g_1g_2g_4g_5$, we can write the equation in (\ref{NewId}); in turn, we showed how the latter implies the identity in (\ref{Final_NewId}).

Finally we should say that we cannot exclude that the identity in (\ref{Final_NewId}) was already known before, however we were not able to find it anywhere in the literature. 

\section{A new identity for the $15j$}\label{4D_section}

\begin{figure}[!ht]
\centering
\includegraphics[width=5cm]{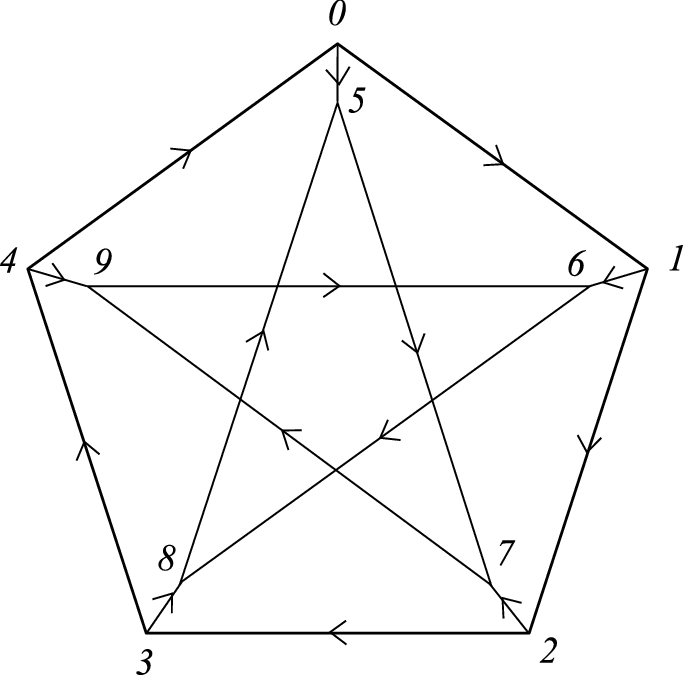}
\caption{\label{specific_4simplex_fig}Our 4-simplex.}
\end{figure}

After showing how the BE identity can be thought as moving a grasping operator along edges of a tetrahedron, we want to extend the concept to four spacetime dimensions, thus deriving a recursion relation for the $15j$-symbol. In order to do that we have to choose a specific $4$-$simplex$ (figure \ref{specific_4simplex_fig}).

We can write the spin-network state for this $4$-$simplex$ explicitly; using the convention $g_{ij}=g_{ji}^{-1} \in G$ 

\begin{eqnarray*}
 s^{ j_{ij}}_{\rm 4-sim}=&  (-1)^{\sum_{i<j}^6(j_{ij}-a_{ij})}\\
& 
\beginpmatrix  j_{01} &j_{05} &j_{04}\\-a_{01} &-a_{05} &b_{04}\pmatrixend
\beginpmatrix  j_{58} &j_{05} &j_{57}\\b_{58} &b_{05}  &-a_{57}\pmatrixend 
\beginpmatrix  j_{12} &j_{16} &j_{01}\\-a_{12} &-a_{16}  &b_{01}\pmatrixend \\ 
&
\beginpmatrix  j_{69} &j_{16} &j_{68}\\b_{69} &b_{68}  &-a_{69}\pmatrixend 
\beginpmatrix  j_{23} &j_{27} &j_{12}\\-a_{23} &-a_{27}  &b_{12}\pmatrixend 
\beginpmatrix  j_{57} &j_{27} &j_{79}\\b_{57} &b_{27}  &-a_{79}\pmatrixend \\
&
\beginpmatrix  j_{34} &j_{38} &j_{23}\\-a_{34} &-a_{38}  &b_{23}\pmatrixend 
\beginpmatrix  j_{68} &j_{38} &j_{58}\\b_{68} &b_{38}  &-a_{58}\pmatrixend 
\beginpmatrix  j_{04} &j_{49} &j_{34}\\-a_{04} &-a_{49}  &b_{34}\pmatrixend\\
& 
\beginpmatrix  j_{79} &j_{49} &j_{69}\\b_{79} &b_{49}  &-a_{69}\pmatrixend
\prod_{i<j} \braketop{j_{ij},a_{ij}}{D^{j_{ij}} (g_{ij})}{j_{ij},b_{ij}}.
\end{eqnarray*}

Now we want to assume flatness on the 10 faces. The flat 4-simplex configuration is given by the analogous of (\ref{phys_tetrahedron}):

\be\label{flat_4simplex}
\psi_{flat}(g_{ij}) = \prod_{f}^{10} \delta(hol_{f}).
\ee

The corresponding $\{15\}$j-symbol is given by:

\begin{eqnarray}\label{Integral_with_deltas_in_4D}
\nn \psi_{15j}(j_{01}\ldots j_{89}) &= \int \prod_{i<j} dg_{ij} \ s^{\prod_{i<j} j_{ij}}_{\rm 4-sim} (\prod_{i<j} g_{ij})\ \psi_{flat}(\prod_{i<j} g_{ij}),\\
&= \{15j\}.
\end{eqnarray}

\begin{figure}[!ht]
\centering
\includegraphics[width=6cm]{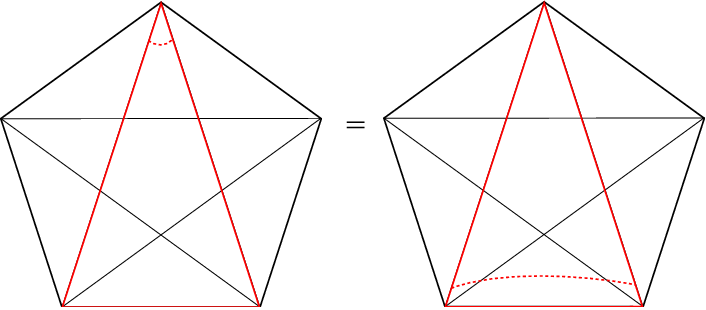}
\caption{The identity diagram.}\label{the_id_diagram}
\end{figure}

As before, to derive an identity we use the fact that, given flatness on a face, the grasping on a vertex is equivalent to the grasping on the other two. We show this identity in figure \ref{the_id_diagram}.

\begin{figure}[!ht]
\centering
\includegraphics[width=5cm]{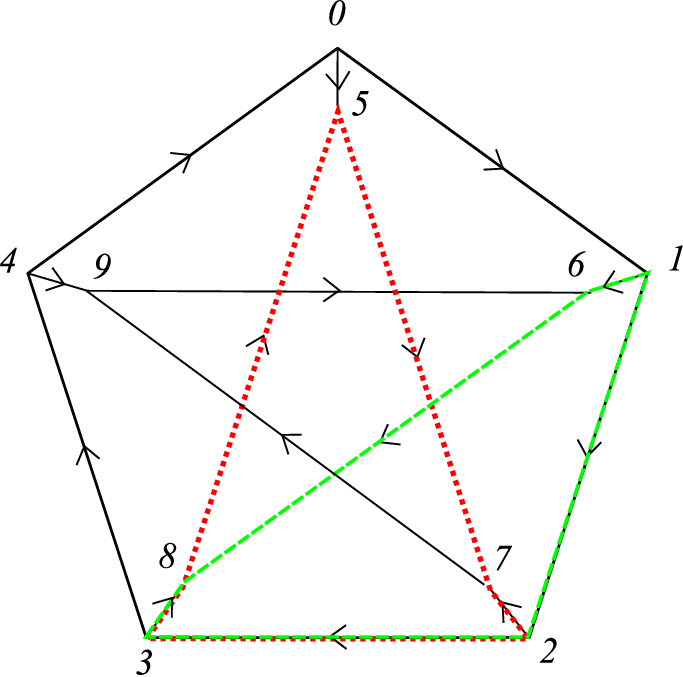}
\caption{\label{RedGreenFaces} Faces of a $15j$.}
\end{figure}

We notice that in our case we can apply our condition to two different ``kinds'' of faces, which are identified by red(dotted) and green(dashed) colouring, as in figure \ref{RedGreenFaces}.

\begin{figure}[!ht]
\centering
\includegraphics[width=7cm]{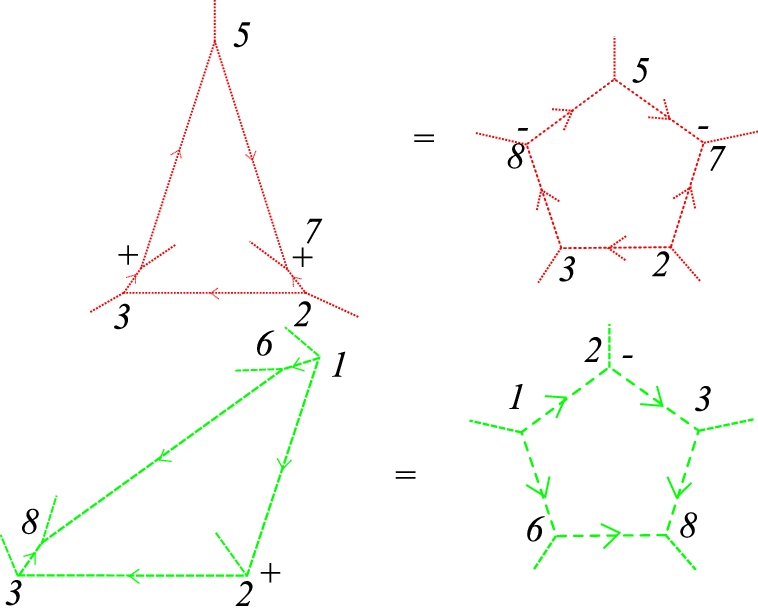}
\caption{\label{RedGreenFaces2} The ``two'' kinds of faces}
\end{figure}

 However, on a closer look, these apparently diverse shapes turn out to be very similar (see fig \ref{RedGreenFaces2}). In the following we will evaluate the new identity for the particular face coloured in red(dotted).

\subsection{The new identity}

\begin{figure}[!ht]
\centering
\includegraphics[width=7cm]{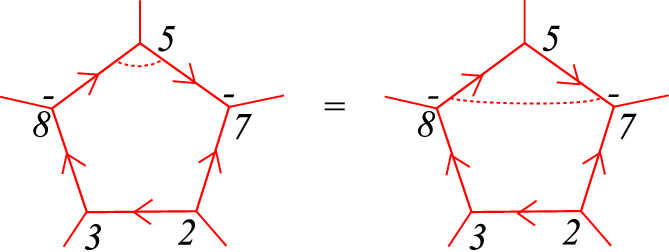}
\caption{\label{diagram_new_id} Diagram of the new identity.}
\end{figure}

 Using the fact that $g_{58} \ g_{83} \ g_{32} \ g_{27} g_{75}=1$, the identity we obtain is schematically the one in figure \ref{diagram_new_id}.

In formulas, we can define a new quantum operator $\hat{H}_{(4D)}$ as following

\be
\fl \hat{H}_{(4D)} s^{ j_{ij}}_{\rm 4-sim}=0, \quad where \quad \hat{H}_{(4D)}=X_{58}\cdot X_{57}-X_{58}\cdot Ad_{hol} X_{57},
\ee

 which yields the identity

\begin{eqnarray}\label{New_4d_id}
\nn X_{58}\cdot X_{57} \; s^{\{j_e\}}_{\rm 4-sim}=X_{58}\cdot Ad(g_{58} \ g_{83} \ g_{32} \ g_{27} \ g_{75}) X_{57} \ s^{\{j_e\}}_{\rm 4-sim}= 
\\
\tl{X}_{58}\cdot Ad( g_{83} \ g_{32} \ g_{27} ) \tl{X}_{57} \ s^{\{j_e\}}_{\rm 4-sim}. 
\end{eqnarray}

Let us split the calculation once again in left and right hand sides and proceed step by step.

\subsubsection{LHS of new $\{15j\}$ identity}

Following an analogous calculation as in section \ref{New_id_section} this part of the calculation is trivial. We get 

\begin{eqnarray}\label{4dId_LHS}
 X_{58}\cdot X_{57} \  s^{\{j_e\}}_{\rm 4-sim}= (-1)^{(j_{58}+j_{57}+j_{05}+1)} N_{j_{58}}N_{j_{57}} \beginBmatrix j_{58} & j_{58} & 1\\ j_{57} & j_{57} & j_{05}\Bmatrixend \ s^{\{j_e\}}_{\rm 4-sim}.
\end{eqnarray}

For the sake of simplicity we will define the amplitude

\be
A_0(j_{01}\ldots j_{89})=(-1)^{(j_{58}+j_{57}+j_{05})}  \beginBmatrix j_{58} & j_{58} & 1\\ j_{57} & j_{57} & j_{05}\Bmatrixend.
\ee

\subsubsection{RHS of new $\{15j\}$ identity}
\begin{figure}[!ht]
\centering
\includegraphics[width=9cm]{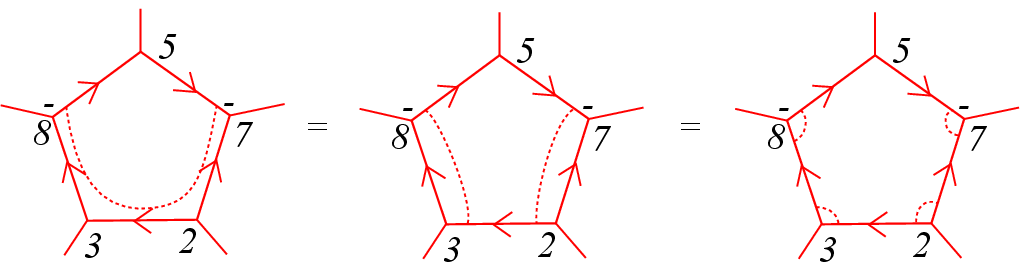}
\caption{Diagrams for the computation of the RHS.}\label{redsteps}
\end{figure}

The right hand side is more complicated. We are going to follow the simple scheme in figure \ref{redsteps}.

The $Ad(g)$ in (\ref{4dId_LHS}) acts as a multiplicative operator in the adjoint representation:

\begin{eqnarray*}
Ad( g_{83} \ g_{32} \ g_{27}) \rightarrow  \ D^1(g_{83})_{m \a}\ D^1(g_{32})_{\a \beta}\ D^1(g_{27})_{\beta n}.
\end{eqnarray*}

First we note that for the representation matrices the following relation holds 

\begin{equation*}
D^j_{mn}(g\I)=(D^j_{nm}(g))^\star=(-1)^{n-m}D^j_{-n-m}(g).
\end{equation*}

Now, as in (\ref{parallel_momenta}), we can compute the products:

\begin{eqnarray}
\fl \nn  (-1)^{(\beta-\a)} \braketop{1,-\beta}{g_{23}}{1,-\a} \braketop{j_{23},a_{23}}{g_{23}}{j_{23},b_{23}}\\
\nn = (-1)^{2j_{23}}\sum_{J_{23}=j_{23}-1}^{j_{23}+1} d_{J_{23}}\ \beginpmatrix 1 & j_{23} & J_{23}\\ \a & -b_{23} & B_{23}\pmatrixend(-1)^{J_{23}-A_{23}}\,\beginpmatrix 1 & j_{23} & J_{23}\\ -\beta & a_{23} & -A_{23}\pmatrixend \\
 (-1)^{1-\beta+j_{23}-b_{23}}\,\langle J_{23},A_{23}\vert g_{32}\vert J_{23}, B_{23}\rangle,
\end{eqnarray}

also

\begin{eqnarray}
\fl \nn (-1)^{(\a-m)}\braketop{1,-\a}{g_{38}}{1,-m}\braketop{j_{23},a_{38}}{g_{38}}{j_{38},b_{38}}\\
\nn = (-1)^{2j_{38}}\sum_{J_{38}=j_{38}-1}^{j_{38}+1} d_{J_{38}}\ \beginpmatrix 1 & j_{38} & J_{38}\\ m & -b_{38} & B_{38}\pmatrixend(-1)^{J_{38}-A_{38}}\,\beginpmatrix 1 & j_{38} & J_{38}\\ -\a & a_{38} & -A_{38}\pmatrixend \\
(-1)^{1-\a+j_{38}-b_{38}}\,\langle J_{38},A_{38}\vert g_{83}\vert J_{38}, B_{38}\rangle,
\end{eqnarray}

and at last

\begin{eqnarray}
\fl \nn \braketop{1,\beta}{g_{27}}{1,n}\braketop{j_{27},a_{27}}{g_{27}}{j_{27},b_{27}}\\
\nn = (-1)^{2j_{27}}\sum_{J_{27}=j_{27}-1}^{j_{27}+1} d_{J_{27}}\ \beginpmatrix 1 & j_{27} & J_{27}\\ \beta & a_{27} & -A_{27}\pmatrixend(-1)^{J_{27}-A_{27}}\,\beginpmatrix 1 & j_{27} & J_{27}\\ -n & -b_{27} & B_{27}\pmatrixend \\
(-1)^{1-n+j_{27}-b_{27}}\,\langle J_{27},A_{27}\vert g_{83}\vert J_{27}, B_{27}\rangle.
\end{eqnarray}

Now I evaluate the action of the operator $L_{m,n}$ on the states:

\begin{eqnarray}
\fl \nn \bra{j_{58}, a_{58}}  L_{m} = (-1)^{2j_{58}+1} N_{j_{58}}\sum_{k_{58}} (-1)^{1-m}(-1)^{j_{58}-k_{58}}\beginpmatrix 1 &j_{58} &j_{58}\\-m &-k_{58} &a_{58}\pmatrixend\,\bra{j_{58}, k_{58}},
\\
\fl    (-1)^{1-n} L_{-n} \ket{j_{57},b_{57}} =  (-1)^{2j_{57}+1} N_{j_{57}}\sum_{k_{57}} (-1)^{j_{57}-b_{57}}\beginpmatrix 1 &j_{57} &j_{57}\\n &-b_{57} & k_{57}\pmatrixend\,\ket{j_{57}, k_{57}},
\end{eqnarray}

The result on the $rhs$ is the following:


\begin{eqnarray}\label{4dId_RHS}
\fl \nn \tl{X}_{58}\cdot Ad( g_{83} \ g_{32} \ g_{27} ) \tl{X}_{57} \ s^{\{j_e\}}_{\rm 4-sim} \\
\fl\nn= \sum_{J_{23}=j_{23}-1}^{j_{23}+1} d_{J_{23}}\ \sum_{J_{38}=j_{38}-1}^{j_{38}+1} d_{J_{38}}\ \sum_{J_{27}=j_{27}-1}^{j_{27}+1} d_{J_{27}}\ 
  N_{j_{58}}\   N_{j_{57}}\ (-1)^{(j_{12}+j_{23}+J_{23} +2J_{27}+2J_{38}+j_{34} +j_{57}+j_{58}+j_{68}+j_{79})}   \\
\fl\nn  \beginBmatrix 1 & J_{38} & j_{38}\\ j_{68} & j_{58} & j_{58}\Bmatrixend 
   \beginBmatrix 1 & J_{38} & j_{38}\\ j_{34} & j_{23} & J_{23}\Bmatrixend
          \beginBmatrix 1 & J_{23} & j_{23}\\ j_{12} & j_{27} & J_{27}\Bmatrixend
      \beginBmatrix 1 & J_{27} & j_{27}\\ j_{79} & j_{57} & j_{57}\Bmatrixend  s^{\{J_{23},J_{27},J_{38},j_e\}}_{\rm 4-sim}. \\
\end{eqnarray}

\subsubsection{The new identity}\label{4dNew_id_section_final}

We evaluated both left and right hand sides of (\ref{New_4d_id}); now we can state the result which is a new identity for the $15j$-symbol. Let us call $A_{(a,b,c)}$ the amplitude 


\begin{eqnarray}
\fl \nn  A_{(a,b,c)}(j_{01} \ldots j_{89})=  d_{j_{23}+ a} \  d_{j_{27}+b} \  d_{j_{38}+c} \ (-1)^{(j_{12}+2j_{23}+a +2j_{27}+2j_{38}+j_{34} +j_{57}+j_{58}+j_{68}+j_{79})} &  \\
\fl \nn  \beginBmatrix 1 & j_{38}+c & j_{38}\\ j_{68} & j_{58} & j_{58}\Bmatrixend 
   \beginBmatrix 1 & j_{38}+c & j_{38}\\ j_{34} & j_{23} & j_{23}+a\Bmatrixend
   \beginBmatrix 1 & j_{23}+a & j_{23}\\ j_{12} & j_{27} & j_{27}+b\Bmatrixend
   \beginBmatrix 1 & j_{27}+b & j_{27}\\ j_{79} & j_{57} & j_{57}\Bmatrixend, \\
\end{eqnarray}
where $(a,b,c)=-1,0,+1$. Then, from (\ref{4dId_LHS}) and (\ref{4dId_RHS}) we get 
%

\begin{eqnarray}
\fl \nn  \hat{H}_{(4D)} \  s^{\{j_e\}}_{\rm 4-sim}=N_{j_{58}}N_{j_{57}} \left( A_0(j_{23},j_{27},j_{38} \ldots j_{89}) \ s^{\{j_e\}}_{\rm 4-sim}+  \right.
\\
\nn  \left.+\sum_{(a,b,c)=(-,-,-)}^{(+,+,+)} A_{a,b,c} (j_{23}+a,j_{27}+b,j_{38}+c,\ldots j_{89}) s^{\{j_{23}+a,j_{27}+b,j_{38}+c,j_e\}}_{\rm 4-sim} \right)=0.\\
\end{eqnarray}

Finally, if we define the linear combination of spin-networks
\be
 \Psi(g_{01} \ldots g_{89})= \sum_{all \ j's} \left[ \prod_{i<j} d_{j_{ij}}  \right]  \  \psi(j_{01} \ldots j_{89}) \ \  s^{\{j_{ij}\}}_{\rm 4-sim}(g_{01} \ldots g_{89}),
\ee
 
then the action of the new operator $\hat{H}_{(4D)} \ \Psi(g_{01} \ldots g_{89})=0$  imposes the following identity for the $15j$-symbol:

\be\label{4DId_final}
\fl \nn A_{0} (j) \ \psi(j_{23},j_{27},j_{38} \ldots j_{89}) + \sum_{(a,b,c)=(-,-,-)}^{(+,+,+)} A_{a,b,c}  (j) \ \psi(j_{23}+a,j_{27}+b,j_{38}+c, \ldots j_{89})=0.\\
\ee

Here we derived only one possible identity for the $15j$-symbol. Obviously we could apply the same method to obtain many more identities just shifting around the grasping operator in different equivalent configurations. For instance we could have chosen to grasp over two different faces as we did for the tetrahedron in sec. \ref{New_id_section}. However the scope of this paper is only to illustrate the method, not to list all possible identities for a particular $nj$-symbol. 

Yet again, we want to stress that the derivation  of the new identity can be considered itself a proof of it. Indeed the fact that we imposed the triviality of the holonomy implied the identity in equation (\ref{New_4d_id}), which, as we showed, implied the resulting identity in (\ref{4DId_final}).

\section{DISCUSSION}

In this paper we described a method to derive recursion relations on  Wigner $nj$-symbols; in particular we derived and proved new identities for the $6j$ and $15j$. Our method consists in a simple and elegant graphical way to describe the action of a certain $grasping$ operator on different nodes of a spin-network. More precisely we assumed flatness on the dual faces of an $n$-simplex spin-network and used such condition to ``move'' around a grasping operator to form equivalent configurations. We then showed that equating these equivalent states implies an identity for the relative $nj$-symbol.

The inspiration for this work came from the paper in \cite{Bonzom:2011hm}. In this paper Freidel et al. define a quantum $Hamiltonian$ operator in the context of $2+1$ spin foam quantum gravity, then prove that a common Angular Momentum identity for the $6j$-symbol called Biedenharn-Elliott identity can be seen a consequence of imposing flatness on the tetrahedron. Such identity plays a key role in three-dimensional Spin-Foam quantum gravity to prove triangulation independence of the theory. 

In section \ref{BE_id_section} we showed how the results of the paper \cite{Bonzom:2011hm} can be  envisioned in a simple and elegant manner through the adoption of our graphical method. We saw that the BE identity is the result of applying the grasping operator on a node of a flat tetrahedron and moving it alongside the edges of the same face. This lead us to the formulation of a $new$ identity, in sec. \ref{New_id_section}, for the $6j$-symbol. To obtain this new identity we grasped two different nodes and moved the grasping along the edges of the different faces.

 Furthermore, in sec. \ref{4D_section} we employed the same graphical method to derive an identity for the $15j$-symbol. Again, imposing flatness on the dual faces, we obtained an identity grasping on a node and sliding the grasping operator along the edges of a face. The result, presented at the end of sec. \ref{4dNew_id_section_final}, is strikingly simple and can be considered the four-dimensional equivalent of the BE identity.

\section{ACKNOWLEDGMENTS}

We would like to thank K. Krasnov, A. Torres Gomez and C. Scarinci for the many stimulating discussions and relevant suggestions.

\appendix

\section{`	Recoupling theory compendium}\label{RecouplingTheoryAppendix}

We present a short summary of the graphical notation we used for angular momentum theory. For further details we refer to \cite{Brink_Satchler-book}.

\begin{itemize}
\item The Wigner $3mj-Symbol$ is represented as a trivalent node. It is assumed that the momenta have to be read in a counter-clockwise manner, unless the node is marked with a $minus$ sign:

\be
\begin{array}{ccc} \includegraphics[height=2cm]{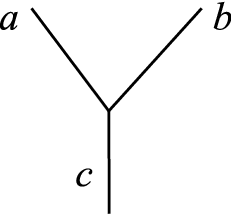} \end{array}= (-1)^{a+b+c}  \begin{array}{ccc} \includegraphics[height=2cm]{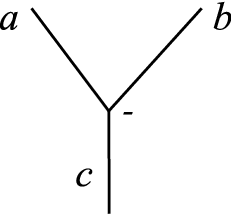} \end{array},
\ee

\be
 \beginpmatrix a &c &b \\ \a &\g &\beta \pmatrixend =(-1)^{a+b+c}\beginpmatrix a &b &c \\ \a &\beta &\g \pmatrixend .
\ee

\item A straight line represents the Kronecker delta

\be\label{Kronecker_delta}
\delta_{ab} \delta_{\a \beta}=  \begin{array}{ccc} \includegraphics[width=1.5cm]{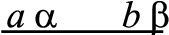} \end{array},
\ee

\item The $Anti$-$symmetric$ $tensor$ is represented as a line marked with an arrow:

\be\label{metric}
(-1)^{a+\a}\delta_{a b} \delta_{\a -\beta}=  \begin{array}{ccc} \includegraphics[width=3cm]{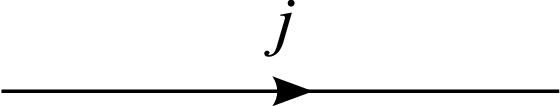} \end{array},
\ee

 Flipping the arrow  produces a phase

\be\label{flipping}
\begin{array}{ccc} \includegraphics[width=3cm]{figure18} \end{array}= (-1)^{2j}  \begin{array}{ccc} \includegraphics[width=3cm]{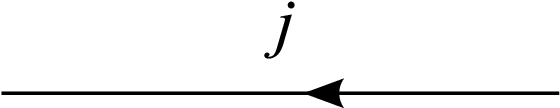} \end{array},
\ee

\item We remind the formula for the {6-j} symbol:

\begin{eqnarray}\label{6j_Symbol}
\fl \nn  \beginBmatrix j_1 &j_2 &j_3\\ j_4 &j_5 &j_6\Bmatrixend = \sum_{a_i,b_i} (-1)^{(j_3+j_2+j_4-a_3 -a_2 -a_4)}
 \\
 \fl\nn \beginpmatrix j_1 &j_3 &j_2\\a_1 &-a_3 &a_2\pmatrixend \beginpmatrix j_3 &j_5 &j_4\\a_3 &a_5 &-a_4\pmatrixend \beginpmatrix j_1 &j_6 &j_5\\a_1 &a_6 &a_5\pmatrixend \beginpmatrix j_2 &j_4 &j_6\\-a_2 &a_4 &a_6\pmatrixend,\\
\end{eqnarray}

\begin{center}
\begin{figure}[!ht]
\centering
\includegraphics[width=4cm]{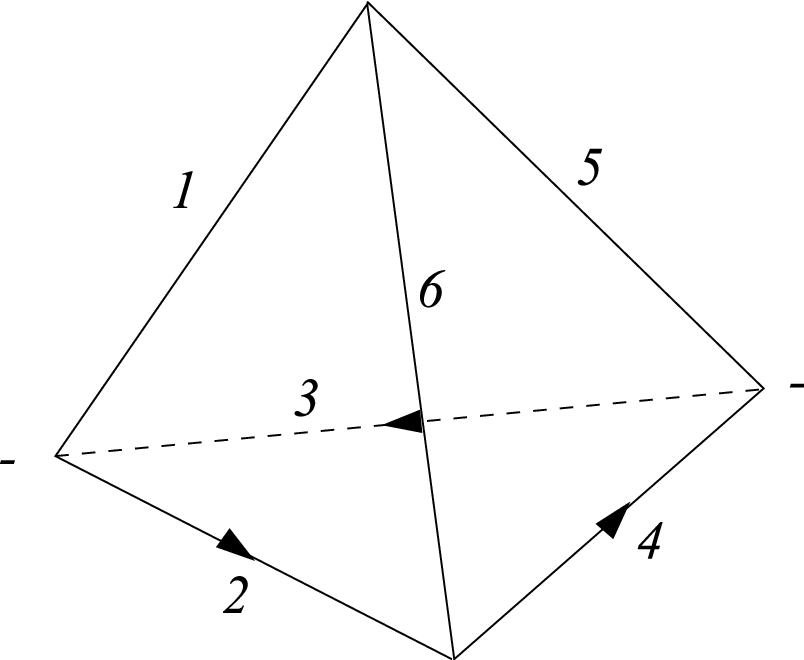}
\caption{\label{recouplingTet} Tetrahedron in graphical notation}
\end{figure}
\end{center}

\item When the internal lines (momenta) of the graph are combined to ``close'' a triangle, one can use the following relation to extract a $\{ 6j\}$ (we follow conventions of \cite{Brink_Satchler-book}):

\begin{figure}[!ht]
\centering
\includegraphics[width=8cm]{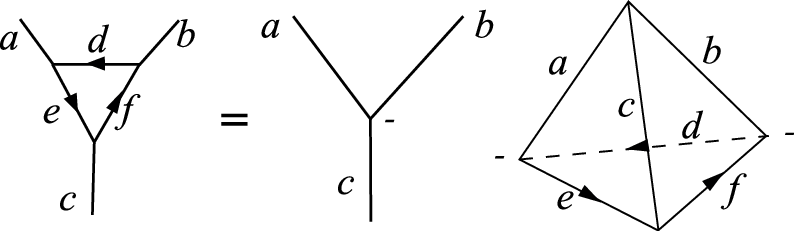}
\caption{\label{graspingbis} Recoupling closed triangle.}
\end{figure}

\begin{eqnarray}\label{grasping}
\nn \sum_{\a,\beta,\g,\delta,\e,\p} (-1)^{d+e+f-\delta -\e-\p} \beginpmatrix e &d &a \\ -\e &\delta &\a \pmatrixend \beginpmatrix f &b &d \\ \p &\beta &-\delta \pmatrixend\beginpmatrix e &c &f \\ \e &\g &-\p \pmatrixend 
 \\
 = \beginBmatrix c &b &a\\ d &e &f\Bmatrixend  \beginpmatrix a &c &b \\ \a &\g &\beta \pmatrixend(-1)^{a+b+c}
\end{eqnarray}

\item Two parallel lines carrying the same group element $g$ can be ``re-coupled'' as in figure \ref{parallel_momenta_figure}
\begin{figure}[!ht]
\centering
\includegraphics[width=9cm]{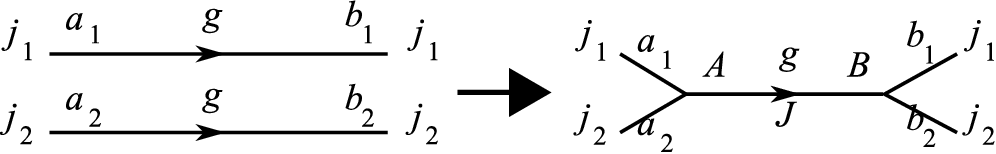}
\caption{\label{parallel_momenta_figure} Recoupling parallel momenta.}
\end{figure}
 
 \begin{eqnarray}\label{parallel_momenta}
\fl \nn \braketop{j_1,a_1}{g}{j_1,b_1} \braketop{1,m}{g}{1,n}=(-1)^{2j_1}\sum_{J_1=|j_1-1|}^{j_1+1} d_{J_1} (-1)^{J_1-A_1} (-1)^{1-n+j_1-b_1}\times
\\
 \nn   \beginpmatrix 1 & j_1 & J_1\\ m & a_1 & -A_1\pmatrixend  \beginpmatrix 1 & j_1 & J_1\\ -n & -b_1 & B_1\pmatrixend
  \langle J,A\vert g\vert J, B\rangle.\\
\end{eqnarray}

\end{itemize}


\end{document}